# INFORMATION AND COMMUNICATION SECURITY MECHANISMS FOR MICROSERVICES-BASED SYSTEMS

Lenin Leines-Vite, Juan Carlos Pérez-Arriaga and Xavier Limón

School of Statistics and Informatics,
Universidad Veracruzana, Xalapa, Ver; Mexico

## ABSTRACT

*Security has become paramount in modern software services as more and more security breaches emerge, impacting final users and organizations alike. Trends like the Microservice Architecture bring new security challenges related to communication, system design, development, and operation. The literature presents a plethora of security-related solutions for microservices-based systems, but the spread of information difficult practitioners' adoption of novel security related solutions. In this study, we aim to present a catalogue and discussion of security solutions based on algorithms, protocols, standards, or implementations; supporting principles or characteristics of information security, considering the three possible states of data, according to the McCumber Cube. Our research follows a Systematic Literature Review, synthesizing the results with a meta-aggregation process. We identified a total of 30 primary studies, yielding 75 security solutions for the communication of microservices.*

## 1. INTRODUCTION

The development of applications based on the microservices architecture has been gaining more and more momentum in enterprise IT [1], this is due to the benefits that the architecture entails, such as low coupling, flexibility, and scalability. This type of software architecture solves the development and scalability problems that were present in monolithic or service-oriented systems (SOA). The microservice architecture bring desirable characteristics: service isolation, functional independence, only responsibility, independent implementation, and light communication [2]. However, the microservice architecture is relatively new, so also new challenges arise in the development of applications based on this type of architecture [3]. These challenges or "pains" as [4] defines them, appear in the design, development, and operation stage of the application.

When integrating an application in a microservices architecture, there are problems related to the communication of its systems, entities, or processes, highlighting confidentiality and integrity issues. Failure to address these issues could compromise the architecture's internal infrastructure, as issues related to confidentiality entail vulnerabilities such as spoofing, illegal access, and replay attacks; and regarding integrity, there are problems such as data interception, manipulation, and leakage [3]. Identifying these problems in the communication of microservices-based systems is important when designing security policies for the development and deployment of the software; it is crucial not to compromise assets and high-value information that are generally exposed on endpoints, and which tend to proliferate. This type of architecture

International Journal of Network Security & Its Applications (IJNSA) Vol.13, No.6, November 2021

tends to be more susceptible, since it requires opening more ports, exposing more APIs, and distributing their access control, thus exposing a more extensive attack surface [4].

This study is a collection of technologies, mechanism and solutions related to confidentiality, authentication, authenticity, integrity, and authorization for the communication of microservices-based systems. In the same way, this study contributes to the understanding of problems in the communication of microservices, highlighting how developers can address them during the development and deployment of the software according with security solutions identified in the Systematic Literature Review (SLR).

This paper is divided into the following sections. Section 2 presents the background and related work, highlighting the common opinions of other authors on the challenges faced by the microservices architecture, and the need to motivate the reinforcement of security in the systems. We address the Systematic Literature Review method in section 3, which considers the review protocol, the results obtained, and the synthesis of the findings. Section 4 presents the discussion of our findings. In Section 5 we conclude the study, setting lines of research and future work.

## 2. BACKGROUND AND RELATED WORK

Microservices-based systems are distributed by nature, they typically run on different hosts or federations of microservices. Each microservice instance is usually a process, therefore, the services can interact with each other using a communication protocol such as HTTP, AMQP, or a binary protocol such as TCP, depending on the nature of each microservice [5].

During the communication of microservices, there are several security problems, as can be seen in Figure 1; so, it is necessary to strengthen the systems internally, since, in its network infrastructure, the services and data contained in the devices connected to the network, are usually very important business and personal assets [6].

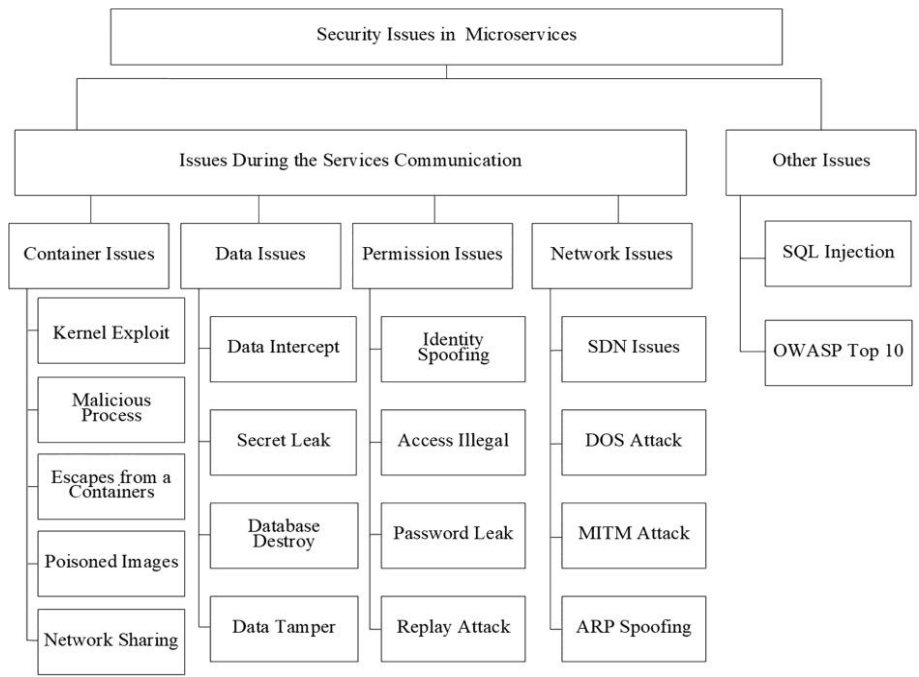

Figure 1. Taxonomy of security problems in microservices.

International Journal of Network Security & Its Applications (IJNSA) Vol.13, No.6, November 2021International Journal of Network Security & Its Applications (IJNSA) Vol.13, No.6, November 2021

**Note.** Adapted from "A survey on security issues in services communication of Microservices-enabled fog applications" by [3], Concurrency and Computation: Practice and Experience, 31. All rights reserved [2018] by John Wiley and Sons. Adapted with permission of the author.

Microservices require more complex communication due to their fine granularity, so [3] highlights that there is not only the risk that the data can be intercepted, but also that malicious entities can infer commercial operations from the shared information. Therefore, [3] mentions that the microservices architecture must verify the authenticity of each service, in addition to verifying the legitimacy of the shared messages, and the valid authorization of origin service. The authors mention that in the study carried out by [7], it is proposed to assign only necessary rights to a subject who requests access to a resource, and that it is valid for the shortest possible time.

The microservices' architecture has had a wide acceptance since it emerged in the industry, therefore, its research has been increasing since then. The literature presents a large number of solutions related to security for the communication of microservices-based systems, but the fact that there isn't a research and compilation on the security methods that can address these problems, makes difficult for practitioners to adopt novel security-related solutions.

We identified four studies that address security for communication in microservices as the main concern:

- [3] performs an analysis on the security vulnerabilities related to the communication of microservices-based systems, considering security problems in four aspects: containers, data, permissions, and network. In addition, the authors mention that, with respect of confidentiality, integrity and availability of the data, consideration should be given to offering minimum data security capabilities, including an encryption scheme, strict access, and storage controls safe. The authors' article addresses the security problems of microservices communication, mentioning resource isolation solutions, container protection, data security, permission security, and network security issues; compared to our article, the solutions exposed in the proposed catalogue present different solution approaches for the communication of microservices, but considering the principles and characteristics of information security [8], the state of the data according to the McCumber Cube [9], and extending the solutions to be used.

- [10] mentions that the problems related to microservices security are multifaceted, so the authors present a security taxonomy in this type of architecture, breaking it down into six categories: hardware, virtualization, cloud, communication, service, and orchestration. His study places microservices and their security in the broader context of SOA and distributed systems, however, the study only considers the security of microservices communication from components deployed in containers, without considering communication with other hardware components, as well as processes and entities. In contrast to our work, we include solutions focused on the protection of the communication of microservices, processes and entities; without considering another layer of security for the architecture, such as hardware protection or virtualization as is considered in the authors' paper.

- [11] aims to provide a useful guide for developers on recognized threats in microservices and how these can be detected, mitigated, or prevented. The study addresses a systematic mapping to discover the main security threats in microservices, introducing an ontology that can serve as a guide for developers to learn about the threats to detect and the security mechanisms to use. Proposing a general security ontology in the microservice architecture leaves research gaps about the parts of the software that developers must

8787

protect. As a result of this, the authors mention the need for studies that intervene in the security of microservices, related to the protection of communication and its individual defence; compared to our study, our study's main objective is to discover solutions that reinforce security for microservices communication and endpoint protection; Therefore, it contributes to the need for studies in the literature that cover these research gaps, as mentioned by the authors.

- [12] addresses a Multivocal Literature Review about the security of microservices-based systems, proposing a classification of security solutions based on the following categories: security solutions, security mechanisms, security scopes, and security contexts. Compared with our study, we propose a catalogue focused on security solutions for the communication of microservices, adding properties to the findings to clarify their nature, defining the security gains of each finding with respect to security problems for the communication of microservices-based systems.

## 3. SYSTEMATIC LITERATURE REVIEW

To guide the study, we followed a Systematic Literature Reviews (SLR) based on the guidelines for performing Systematic Literature Reviews in Software Engineering proposed by [13], which includes planning the review, conducting the review, and data synthesis.

[13] mentions that the guidelines for performing SLR in software engineering are used to conduct rigorous reviews of the current empirical evidence in the software engineering community. It is relevant for our work to include an SLR, since the SLR allows us to use a well-defined methodology to identify, analyse and interpret all the available evidence related to a specific research question. That in the case of this study is appropriate for the investigation, documentation and classification of mechanisms, to help reinforce the confidentiality, authentication, authenticity, integrity, and authorization in the communication of microservices-based systems, in order to provide a catalogue that allows to recognize solutions to guide a secure integration of applications based on this type of architecture; since, as has been mentioned, there are security gaps in the communication of microservices that could allow the exploitation of vulnerabilities and information leakage within the infrastructure, which could be used as a guide for the total control of the system [4].

The exhibition of these solutions is intended to serve as a line of research and application in development projects that lack security frameworks in software, and that both the academic and professional fields can consult as a catalogue of references to guide an environment secure in the interaction of microservices during their deployment.

### 3.1. Planning the review

The objective of the SLR is to analyse solutions related to the principles and characteristics of information security [8], in conjunction with the 2nd dimension of the McCumber Cube [9], concerning the communication of systems, entities and processes built in a microservices-based systems, with the purpose to catalogue and expose the security methods discovered given their similarities.

Initially [13] mentions carrying out a planning of the review, raising research questions as a separator of doubts; the proposed SLR considers three research questions about communication security for microservices-based systems, with the objective of identifying solutions that work to mitigate problems related to confidentiality and integrity, as well as discovering communication protocols used within this context, revealing security mechanisms.

- **Q1.** What security mechanisms are related to the confidentiality of communication in the microservices-based systems?
- **Q2.** What communication mechanisms for communication integrity between microservices are reported in the literature?
- **Q3.** What communication protocols are used in the context of the microservice architecture?

### 3.2. Conducting the review

Table 1. Keywords and related concepts used in the search query

| Keyword | Related concepts |
| --- | --- |
| Microservices | Microservice, micro-service |
| Mechanism | Mechanism, algorithm, protocol, standard, framework |
| Communication | Communication, interaction, connection |
| Authentication | Authentication |
| Authorization | Authorization, consent, permit, permission |
| Confidentiality | Confidentiality, secret |
| Integrity | Integrity, wholeness |
| Information | Information, data |
| Request | Petition, request |

- **Q1.** ("Microservice" OR "Micro-service") AND ("Mechanism" OR "algorithm" OR "standard" OR "framework") AND ("confidentiality" OR "authorization" OR "Authentication") AND ("information" OR "data") AND ("communication" OR "interaction" OR "connection")
- **Q2.** ("Microservice" OR "Micro-service") AND ("Communication" OR "interaction" OR "connection") AND ("mechanism" OR "algorithm" OR "technology" OR "standard" OR "framework") AND ("integrity" OR "wholeness")
- **Q3.** ("Microservice" OR "Micro-service") AND ("protocol") AND ("communication" OR "interaction" OR "connection")

The review conduction considers carrying out a search strategy establishing keywords, as shown in Table 1, to later formulate search strings for determining information sources and databases for the collection of articles. The search process for the SLR was an automated search of conference proceedings and articles from information sources and databases. The sources of information consulted were Emerald, Science Direct, Springer Link, Editorial Wiley, ProQuest, ACM Digital Library, IEEEXplore Digital Library, and Google Scholar.

As [13] suggests, we established a systematic search strategy with inclusion and exclusion criteria to identify the most relevant studies in the literature. Table 2 summarizes the criteria applied in the SLR, presented as selection filters.

#### 3.2.1. Study Selection

We carried out the collection of studies with the search strings in the aforementioned information sources and databases, we replicated each search string in each source, and we applied the filtering stages of Table 2. During the first collection of studies, we identified 17 primary studies.

To enrich the collection of studies, we carried out a second search, conjugating the search strings to cover the maximum number of articles, managing to increase the collection to 30 studies in total between the first and second collection, these studies are in Table 3.

Table 2. Filters for the selection of studies

| Filters | Criteria |
|---|---|
| Without filters | Exclusion criteria are not applied. |
| 1st Filter | Publication date <5 years old from 2020. English language. Publication: congresses, conferences, journals. |
| 2nd Filter | Title: At least one keyword answers the research question. |
| 3rd Filter | Context: the keywords in the abstract or in the conclusion respond to of the research questions directly or indirectly. |
| 4th Filter | Quick reading to confirm the relationship of the study with the question of investigation. |

Table 3. Selected studies

| No. | Studies |
|---|---|
| 1 | Defense-in-depth and Role Authentication for Microservice Systems. [14] |
| 2 | A Cluster of CP-ABE Microservices for VANET. [15] |
| 3 | A Web Service Security Governance Approach Based on Dedicated Micro-services. [16] |
| 4 | A survey on security issues in services communication of Microservices-enabled fog applications. [3] |
| 5 | Capabilities for Cross-Layer Micro-Service Security. [17] |
| 6 | Mechanisms for Mutual Attested Microservice Communication. [18] |
| 7 | Towards Automated Inter-Service Authorization for Microservice Applications. [19] |
| 8 | eZTrust: Network-Independent Zero-Trust Perimeterization for Microservices. [20] |
| 9 | An optimized control access mechanism based on micro-service architecture. [2] |
| 10 | Authentication and authorization orchestrator for microservice-based software architectures. [21] |
| 11 | Towards Multi-party Policy-based Access Control in Federations of Cloud and Edge Microservices. [22] |
| 12 | Graph-based IoT microservice security. [23] |
| 13 | Overcoming Security Challenges in Microservice Architectures. [10] |
| 14 | Identity and Access Control for micro-services based 5G NFV platforms. [24] |
| 15 | DNS/DANE Collision-Based Distributed and Dynamic Authentication for Microservices in IoT. [25] |
| 16 | Applying Spring Security Framework and OAuth2 To Protect Microservice Architecture API. [26] |
| 17 | Design of a micro-service based Data Pool for device integration to speed up digitalization. [27] |
| 18 | Hybrid Blockchain-Enabled Secure Microservices Fabric for Decentralized Multi-Domain Avionics Systems. [28] |
| 19 | Microservice Security Agent Based On API Gateway in Edge Computing. [29] |
| 20 | Secure end-to-end processing of smart metering data. [30] |
| 21 | Secure Cloud Processing for Smart Meters Using Intel SGX. [31] |
| 22 | BlendSM-DDM: BLockchain-ENabled Secure Microservices for Decentralized Data Marketplaces. [32] |
| 23 | Performance Analysis of RESTful API and RabbitMQ for Microservice Web |
| | Application. [33] |
| 24 | A platform-independent communication framework for the simplified development of shop-floor applications as microservice components. [34] |
| 25 | Component-Based Refinement and Verification of Information-Flow Security Policies for Cyber-Physical Microservice Architectures. [35] |
| 26 | Design and implementation of a decentralized message bus for microservices. [36] |
| 27 | Interface Quality Patterns: Communicating and Improving the Quality of Microservices APIs. [37] |
| 28 | Securing IoT microservices with certificates. [38] |
| 29 | Implementing a Microservices System with Blockchain Smart Contracts. [39] |
| 30 | Building Critical Applications Using Microservices. [40] |

### 3.2.2. Data Extraction

For data extraction, we used meta-aggregation as a qualitative synthesis method. This method makes it possible to synthesize the results by aggregating findings into categories [41], easing the grouping of findings according to attributes or properties in common.

To carry out the meta-aggregation process, the JBI (Joanna Briggs Institute) evidence synthesis manual was partially considered, which involves the extraction of findings, which are then grouped into categories. We combined the categories to construct synthesis statements, taking the findings as conclusions reached by researchers, often presented as topics [42]. The process for meta-aggregation is divided into three stages as mentioned by [43]:

- Extract the findings.
  The first step in this process involves the reviewers extracting all the findings from each of the included articles and defining one illustration per finding. A finding is defined as a topic, category, or metaphor reported by the authors of original articles.
- Categorize the findings.
  The second step in the meta-aggregation involves an evaluation of the similarity in the meaning of the findings, which cross the different original articles.
- Synthesize categories.
  The reviewer should review the full list of categories developed and identify sufficient similarity in meaning to generate a complete set of synthesized findings.

As the first step in meta-aggregation, we classified each article in a table, exposing the identified findings and associating the following properties per finding:

- Security model as a reference for security concepts (CIA as the default security model).
- Related layer of the OSI model.
- STRIDE threat model classification [44].
- Security concept based on the principles and characteristics of information security [8]. Regarding the STRIDE model [44], we considered the following security properties mapped with corresponding threats:
  - Confidentiality = Information disclosure
  - Authentication = Spoofing
  - Authorization = Elevation of privilege
  - Authenticity = Spoofing and Repudiation
  - Integrity = Tampering).
- Security technique (s) associated with the discovered algorithm, protocol, standard, or implementation (possible finding ID).

- Name of the discovered algorithm, protocol, standard, or implementation (finding ID).

Subsequently, we categorize each finding according to their related security property and their appearance frequency in the literature, as shown in Table 4. Finally, we catalogued each finding given its nature (protocol, algorithm, standard or implementation), in addition to categorizing the finding according to the 2nd dimension of the McCumber Cube [9], concerning the state in which data security is carried out (transmission, storage, process). This catalogue is shown in Table 5.

Table 4. Grouping of findings in relation to the concept of security

| Security solution | Safety concept | Frequency of mention in the literature | |
|---|---|---|---|
| | | Methods proposed in the studies | Methods mentioned in studies |
| Finding ID (name of discovered algorithm, protocol, standard, or implementation). | • Confidentiality.<br>• Authentication.<br>• Authenticity.<br>• Authorization.<br>• Integrity. | Total mentions of solutions as proposed by the authors. | Total mentions of solutions addressed by the authors and techniques mostly used by developers or autonomous DevOps teams. |

As support for the classification of security solutions in our study, we took as a reference the principles and characteristics of information security [8], the STRIDE threat taxonomy [14], and the McCumber Cube security model. The mentioned models provide the following:

- The CIA triad provides a base model for discussing ideas related to security [45][46]; and is the basis for extended models such as the information security features mentioned by [8], which also addresses current security issues, such as authentication, authorization, and non-repudiation.

- The McCumber cube provides an abstract model for information systems protection. It considered the CIA triad expanding it into three dimensions: Critical Information Characteristics, Information States and Security Measures. More importantly for our study, this model considers the state of the data: storage, transmission, or process; and also security measures: technology, policies, and human factors [9]. By using a multidimensional approach, the McCumber cube provides a complete view of data security, considering important aspects about factors that impact data security. For example, [47] mentions that confidentiality can be considered with respect to data in a state of transmission, storage or process; such data could be encrypted following three different approaches: 1) encrypted data at rest (file encryption), 2) encrypted data in transit (TLS encryption for data on the network), and 3) encrypted data in process (SGX encryption for in-memory data processing).

- STRIDE is a threat modelling approach that identifies threats and risk reduction strategies. It considers threats from the attacker's perspective, thus establishing itself as an opposed model to the CIA triad and information security characteristics [8], [44], [48].

International Journal of Network Security & Its Applications (IJNSA) Vol.13, No.6, November 2021

### 3.2.3. Results

The results of the frequency of the findings according to the abstraction of their nature can be seen in Figure 2, while the result of the frequency of the findings about the principles and characteristics of information security can be seen in Figure 3.

In figure 2 the solutions catalogued as indistinct are findings of different kinds, concerning ideas that are not distinguished as algorithms, protocols, standards or implementations. However, indistinct findings may be related to other findings, for example, the password security method, can be seen as a required parameter to validate an entity, or the software component that is granted privileges.

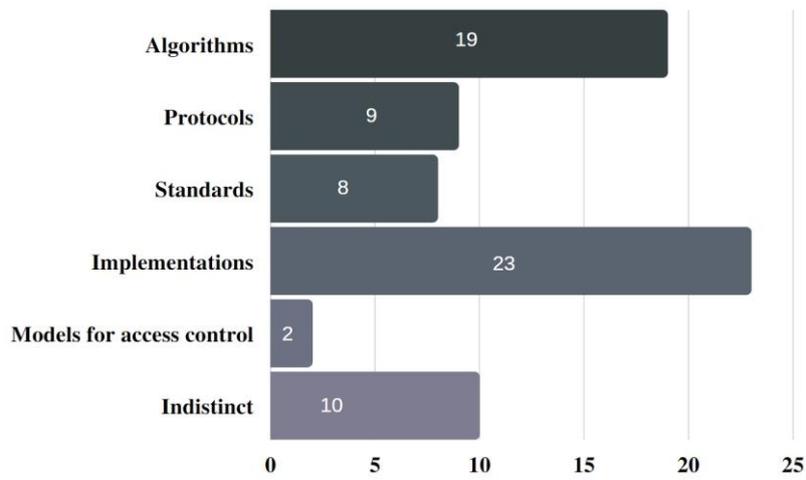

Figure 2. Frequency by category of the findings according to their nature

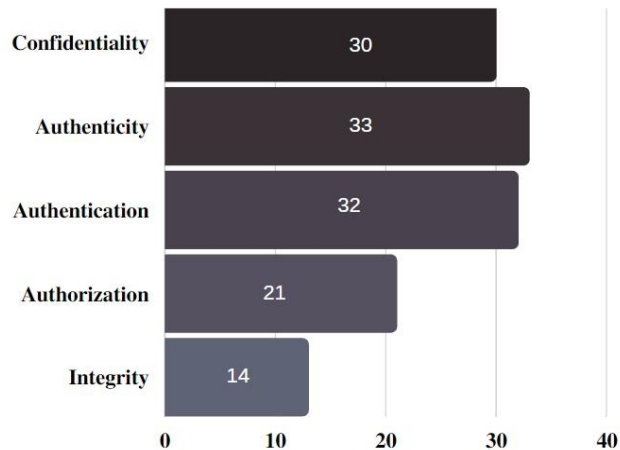

Figure 3. Frequency by category of the findings in relation to the principles and characteristics of information security

## 3.3. Data Synthesis

The diagram in Figure 4 shows the 71 findings identified and their frequency in the literature. We grouped the results according to their security property; we also considered the existing relationships between the security concepts presented by the findings. The synthesis of the findings is outlined in two colour tones, blue and red, as there are two classes. The first class are the methods proposed by the authors, identified by blue tones; while the second class are methods mentioned by the authors, identified by red tones. The grouping for the findings related to the concept of confidentiality involved for the most part heterogeneity with other concepts of security, for this reason, the findings located in the center of the diagram in Figure 4 are groupings related to the concept of confidentiality, while the findings that are on the edge of the diagram, are groupings that are directly associated with their concept. Also, the relationship with other security concepts can exists, we make this distinction using labels in the diagram.

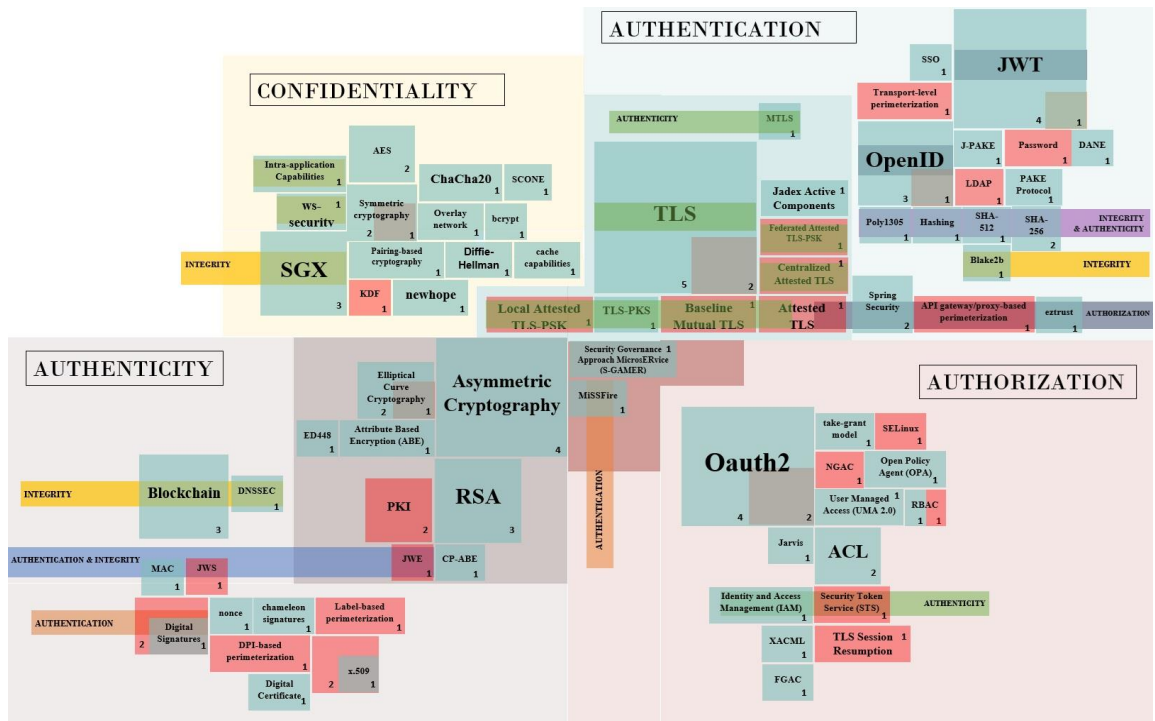

Figure 4. Diagram of findings identified in the literature through meta-aggregation

## 3.4. Threats to the validity of the study

This subsection poses threats to the validity of the proposed SLR, since, regarding the criteria of the Guidelines for performing Systematic Literature Reviews in Software Engineering proposed by [13], the criterion for evaluating the quality of the papers to be included in the SLR, it was not considered. We decided to exclude this criterion, since the sources consulted from where the primary articles were extracted are reliable sources with a track record in computer science, or in various academic areas, so the quality of the studies is high. However, during the discussion of the findings, the lack of discussion of quality metrics of each study included in the SLR was noted, assisted by criteria aimed at questioning their veracity, performance, efficiency, among other critical properties. Therefore, it is proposed to address this issue as future work, and it is discussed in the final part of the study.

## 4. DISCUSSION

In a network infrastructure, there are three categories of network components: Devices, Media, and Services [16]. In microservices architecture, these components can be abstracted since this type of architecture allows interactions within the same system or cluster of microservices.

Therefore, the use of security techniques is different from those used in a normal network infrastructure, as the medium is not only communication channels, but it also involves interconnections between processes of microservices, which increases the number of inter-process communications, the number of context switches, and the number of I/O operations [49]. Given the diversity of approaches that we found of security solutions for microservices communication in the literature, we present the following classification:

- Security methods aimed at reinforcing security in some layers of the OSI model, for data security, as well as providing Defence in Depth or considering a Zero Trust building approach, as recognized in the literature. Possibly considering a security model, such as the McCumber Cube [9], the STRIDE threat taxonomy [44], or the principles and characteristics of information security [8].
- Solutions aimed at a specific microservices communication problem. For example, in IoT, VANET networks, NFV (Network Function Virtualization), among others.
- Solutions based on existing security methods.
- Solutions proposed by the authors.
- Solutions oriented to a computing paradigm. For example, Cloud Computing or Edge Computing.
- Solutions focused on data security.
- Solutions focused on the security of the communication of entities or software components.

After having synthesized the findings through meta-aggregation, it was possible to answer the research questions. Table 5 presents a catalogue divided by categories according to the nature of the finding. These solutions represented as findings are susceptible to the implementation context, so their approach is important. To clarify the differences between these solutions, we made a distinction according to the state of the data, or if the security approach is associated with security for the interaction between systems and entities. In relation to **Q1** we found that all the solutions are associated with the principle of confidentiality, or with the information security characteristics (authenticity, authentication, or authorization).

The analysis of these 30 studies served as the basis to understand the security problems faced by microservices communication, and how the solutions presented can address these protection needs for the system infrastructure. Of the 71 findings identified in the literature, the authors mention them with different frequency, highlighting:

- The SGX algorithm, used to provide a Trusted Execution Environment, protecting memory regions, data, and encryption keys [50]. For example, [40] uses SGX in conjunction with secure containers for deploying microservices, protecting in this way the CPU registers, the Main memory, system files, and network communication. This mechanism preserves confidentiality and integrity, even against higher privilege attacks targeted to the operating system or the hypervisor.

- The TLS protocol, mentioned in 8 studies, in addition to finding 7 implementations based on TLS. The protocol is a layer of confidentiality and integrity for the communication channel between pairs of components of software in a microservices architecture. According to [14], most microservices systems employ HTTP-based approaches, so TLS communication is generally used to protect communication channels as part of HTTPS; and in agreement with the authors, [25], mention that IoT/M2M protocols generally rely on TLS for authentication and end-to-end protection in microservices-based systems. The literature presents variants of TLS, for example, [10] mentions the use of Mutual

Transport Layer Security (MTLS) with a self-hosted public key infrastructure (PKI) as a method to protect all internal service to service communications, pointing out two different examples to establish trust between microservices: The Docker Swarm Case, and the Netflix case. MTLS is also mentioned in [18], proposing mutual authentication mechanisms for microservices by coupling MTLS with platform certifications, from a Trusted Platform Module (TPM).

- The JSON Web Token (JWT) standard, mentioned in 5 studies, and also OAuth 2.0, mentioned in 6 studies. These security standards consider authorization between systems in a network environment. [10] discusses token-based authentication schemes for microservices that allow user-to-service authentication and identity propagation based on cryptographic tokens, such as JWT. On the other hand, [3], [22], [24], and [26], propose OAuth 2.0 as a viable solution for the authorization of access to systems or resources for microservices.

We also found interesting security mechanisms:

- [18] mention seven different mechanisms based on TLS as a transport protocol, and Tao services, a trusted cloud platform proposed by [51]. The mechanisms include symmetric, asymmetric, or shared keys, for mutual authentication between microservices.

- [20] proposes eZTrust, a model oriented to Zero Trust security, proposed as an alternative solution for security perimeters of network endpoints, considering workload identities. eZTrust enables data center tenants to express access control policies based on detailed workload identities and enables data center operators to enforce those policies reliably and efficiently, independently of the network. eZTrust reuses data monitoring obtained from deployed workloads for edge security, each packet generated by a microservice is marked with a tag that encodes a detailed identity of the microservice, and the detailed identity is defined as a set of authentic contexts linked to the workload of the microservice. The study shows, step by step, how the discovery of the context, the labeling and the verification of packages was carried out by a program based on the Berkeley Extended Packet Filter (eBPF). This allows active probing through a collector, which is an infrastructure privileged monitoring service, collecting additional microservices contexts related to the runtime environment, either by querying the centralized microservices orchestrator or by connecting itself to the namespaces of the target microservice.

- [19] mentions that authorization mechanisms between distributed systems usually use complex access control policies to perform detailed authorization, and due to the high number of systems that are commonly presented in microservices applications, it is unrealistic for administrators to manually configure and maintain policies for the access control. Therefore, for an access control mechanism to work, automation is essential within this context. [19] proposes in their study to Jarvis, an automated authorization mechanism between services for microservices applications. The framework, through static analysis, automatically extracts the possible invocations that a microservice can initiate, and then generates detailed access control policies, considering all the information that reflects the invocation relationships between microservices. In addition to monitoring microservices' changes and quickly adjusting access control policies.

Concerning Q2, we identified 14 solutions related to information integrity. In the same way, we found interesting security solutions:

- Concerning data integrity attack mitigation, [17] mentions verifying at runtime data integrity of high-value assets, such as databases and reverse proxies. A practical example an attack is sending authentication credentials to a web server, an attacker can unexpectedly manage to divert the flow of control within the web server to capture the data, and thus control the integrity of the flow (CFI), which could not manifest any change at the system level, going unnoticed in the application. It is useful to verify incorrect configurations and internal threats, so [17] generates through a C abstract machine, instructions for memory access capacity lists, this can augment CFI methods, in addition to mitigating Control Flow Bending and Control Jujutso attacks. [17] takes advantage of a data provenance-based approach to produce instructions for lists of memory access capabilities. Data provenance requires tagging all memory writes, tracking memory access during program execution, and finally stopping execution when access deviates from a reliable data provenance graph for a given buffer.

- The implementation of Smart Contracts is also a novel security-related solution, these are programs that live and run on a blockchain backbone; written for the purpose of enforcing agreements between two parties in a decentralized and unreliable environment without the control of a central authority. [39] aims to demonstrate that it is possible to fully implement a microservices-based systems with Smart Contracts, taking advantage of the principle that each change and operation is permanently and transparently recorded in the blockchain ledger. Since, in general, microservices can be mapped in an architecture, where smart contracts and blockchain are the backbone of the software services provided.

- Hash functions: SHA-256, SHA-512 and Blake2b, used for digital signatures, integrity protection mechanisms, as well as authentication and authenticity for data at transit. These hashing algorithms deal with data integrity checks in the communication of microservices [14], [31], [38].

- Specific algorithms for message authentication, such as Poly1305, contained in the cryptosuite of the implementation proposed by [14], or the handling of message authentication codes (MAC).

Table 5. Catalogue of security solutions for microservices communication

**NA (Not Applicable).** It is because the method is not involved with information security but with the communication of the entities, services or processes that communicate in the microservices architecture.

| | Security method | Safety concept | | | | | Security applied according to the state of the data | | |
|---|---|---|---|---|---|---|---|---|---|
| Algorithms | Intra-application capabilities | X | ° | ° | ° | X | NA(X) | | |
| | Cache capabilities | X | ° | ° | ° | ° | ° | ° | X |
| | New Hope | X | ° | ° | ° | ° | NA | | |
| | ChaCha20 | X | ° | ° | ° | ° | X | X | ° |
| | Bcrypt | X | ° | ° | ° | ° | X | X | ° |
| | Hashing | ° | X | X | ° | X | X | X | ° |
| | SHA-256 | ° | X | X | ° | X | X | X | ° |
| | SHA-512 | ° | X | X | ° | X | X | X | ° |
|  | RSA | X | ° | X | ° | ° | X | X | ° |
|  | Poly1305 | ° | X | X | ° | X | X | X | ° |
|  | Blake2b | ° | X | ° | ° | X | X | X | ° |
|  | MAC | ° | X | X | ° | X | X | ° | ° |
|  | Chameleon Signatures | ° | ° | X | ° | ° | X | ° | ° |
|  | CP-ABE | X | ° | X | ° | ° | X | X | ° |
|  | Attribute-based encryption (ABE) | X | ° | X | ° | ° | X | X | ° |
|  | ED448 | X | ° | X | ° | ° | X | ° | ° |
|  | AES | X | ° | ° | ° | ° | X | X | ° |
|  | KDF | X | ° | ° | ° | ° | NA | | |
|  | SGX | X | ° | ° | ° | X | ° | ° | X |
| **Protocols** | TLS | X | X | X | ° | ° | X | ° | ° |
|  | TLS-PKS | X | X | X | ° | ° | X | ° | ° |
|  | Mutual authentication TLS (MTLS) | X | X | X | ° | ° | X | ° | ° |
|  | Diffie-Hellman | X | ° | ° | ° | ° | NA(X) | | |
|  | WS-Security | X | ° | ° | ° | X | X | ° | ° |
|  | LDAP | ° | X | ° | ° | ° | NA | | |
|  | PAKE Protocol | ° | X | ° | ° | ° | X | ° | ° |
|  | J-PAKE | ° | X | ° | ° | ° | X | ° | ° |
|  | User Managed Access (UMA 2.0) | ° | ° | ° | X | ° | NA | | |
| **Standards** | JWT | ° | X | ° | X | ° | NA(X) | | |
|  | JWE | X | X | X | ° | X | X | ° | ° |
|  | JWS | ° | X | X | ° | X | X | ° | ° |
|  | Oauth 2.0 | ° | ° | ° | X | ° | NA | | |
|  | OpenID | ° | X | ° | X | ° | NA | | |
|  | Security token service (STS) | ° | ° | X | X | ° | NA | | |
|  | DANE | ° | X | ° | ° | ° | NA | | |
|  | x.509 | ° | ° | X | ° | ° | NA | | |
| **Implementations** | Security Governance Approach MicrosERvice | X | ° | ° | X | ° | NA | | |
|  | Attested TLS | X | X | X | X | ° | X | ° | ° |
|  | Centralized Attested TLS | X | X | X | ° | ° | X | ° | ° |
|  | Centralized Attested TLS-PSK | X | X | X | ° | ° | X | ° | ° |
|  | TLS Session Resumption | ° | ° | ° | X | ° | X | ° | ° |
|  | Local Attested TLS-PSK | X | X | X | ° | ° | X | ° | ° |
|  | Federated Attested TLS-PSK | X | X | X | ° | ° | X | ° | ° |
|  | Baseline Mutual TLS | X | X | X | ° | ° | X | ° | ° |
|  | Jadex Active Components | X | X | ° | ° | ° | NA | | |
|  | API gateway/proxy-based perimeterization | ° | X | ° | X | ° | NA | | |
|  | eztrust | ° | X | ° | X | ° | NA | | |
|  | Open Policy Agent (OPA) | ° | ° | ° | X | ° | NA | | |
|  | Spring Security | ° | X | ° | X | ° | NA | | |
|  | SCONE | X | ° | ° | ° | ° | ° | X | ° |
|  | Transport-level perimeterization | ° | X | X | ° | ° | NA | | |
|  | Label-based perimeterization | ° | ° | X | ° | ° | NA | | |
|  | DPI-based perimeterization | ° | ° | X | ° | ° | NA | | |

| | | | | | | | | | | |
|---|---|---|---|---|---|---|---|---|---|---|
| | Jarvis | ○ | ○ | ○ | X | ○ | | NA | | |
| | blockchain | ○ | ○ | X | ○ | X | ○ | | X | ○ |
| | DNSSEC | ○ | ○ | X | ○ | X | | NA(X) | | |
| | PKI | X | X | X | ○ | ○ | | NA | | |
| | MiSSFire | X | X | ○ | X | ○ | X | | ○ | ○ |
| | SELinux | ○ | ○ | ○ | X | ○ | | NA (X) | | |
| **Models for access control** | Take-grant model | ○ | ○ | ○ | X | ○ | | NA | | |
| | RBAC | ○ | ○ | ○ | X | ○ | | NA | | |
| **Indistinct solutions** | Password | ○ | X | ○ | ○ | ○ | | NA | | |
| | ACL | ○ | ○ | ○ | X | ○ | | NA | | |
| | Identity and Access Management | ○ | ○ | X | X | ○ | | NA | | |
| | Overlapping network | X | ○ | ○ | ○ | ○ | X | | ○ | ○ |
| | XACML | ○ | ○ | ○ | X | ○ | | NA | | |
| | NGAC | ○ | ○ | ○ | X | ○ | | NA | | |
| | SSO | ○ | X | ○ | ○ | ○ | | NA | | |
| | Digital certificate | ○ | ○ | X | ○ | ○ | | NA | | |
| | Digital signature | ○ | ○ | X | ○ | X | X | | ○ | ○ |
| | nonce | ○ | ○ | X | ○ | ○ | X | | ○ | ○ |

■ **Confidentiality** ■ **Authentication** ■ **Authenticity** ■ **Authorization** ■ **Integrity**

■ **Transmission** ■ **Storage** ■ **Process**

Finally, in response to **Q3**, we identified four protocols to establish secure communication, and ten protocols only oriented to the communication of systems. The findings include cryptographic protocols such as Diffie-Hellman and TLS; protocols for secure communication, such as gRPC and WS-Security; communication and message protocols such as RESTful, RabbitMQ, AMQP, ZeroMQ, Google Protobuf serializer, OPC Unified Architecture (OPC UA), Extendable Machine Connector (XSC), Pasty Protocol; and, Pastry or Scribe, as protocols for the discovery of services.

In addition, we found six articles that discuss protocols and API construction patterns for the communication of systems, entities, and processes in a microservices architecture:

- [29] discusses edge computing platforms, providing customers with a secure API Gateway, which is a microservice security agent based on the API Gateway Kong (https://konghq.com/). This solution can run in front of any REST API and can be extended by plugins such as JSON Web Token (JWT) authentication.

- [33] presents a performance comparison between RESTful API and RabbitMQ for web applications based on microservices architectures from the results, the authors conclude that with few users (50-200 users), RabbitMQ shows a slow response time compared to REST API. However, with 250 users, the number of requests exceeds the processing load, and the REST API performance gradually degrades, errors that lead to service loss can also occur.

- [34] proposes the use of microservices within industry 4.0 production processes. For evaluation, the study proposes three different communication implementations: OPC UA (OPC Foundation, 2008), XSC (RetroNet research project of the Institute of Technology for the Control of Machine Tools and Manufacturing Facilities of the University of Stuttgart; ISW), and Beckhoff ADS (Automation Device Specification), plus two-hybrid configurations: Beckhoff ADS over XSC (ADS-XSC) and OPC UA over XSC (OPC-XSC). The results show that XSC outperforms the two protocols, ADS and OPC UA, for small packets of up to 5120 bytes, which is why they mention the viability of XSC within industry 4.0 production processes.

- [35] addresses the refinement and verification of information flow security policies for cyber-physical microservices architectures, based on MechatronicUML [52]. This provides a set of rules to refine a macro-level policy to a micro-level, for highly granular systems, such as microservices. In addition to this, the study presents a safety verification technique for real time microservices's message passing, avoiding information leaks concerning the composition of the microservices architecture.

- [36] mentions the need of communication decoupling for the microservices architecture. The study proposes a microservices design using a decentralized message bus to use as a communication tool between services. The study also proves a framework for service collaboration, divided into four main components: public API, message bus, messaging, and service discovery.

- [37] proposes quality patterns for improving the design and use of APIs in the microservices architecture. The study basically describes five patterns: API Key, Wish List, Rate Limit, Rate Plan, and Service Level Agreement. This approach focuses on the quality characteristics of an API, as well as its negotiation and agreement.

## 5. CONCLUSIONS

With the help of the Guidelines for performing Systematic Literature Reviews in Software Engineering proposed by [13], and the meta-aggregation process [41], we achieved the objective of this study. We identified solutions related to the principles and characteristics of information security [8]. We grouped the solutions discovered based on their properties. We identified protocols and API construction patterns for the communication of systems, entities, and processes in a microservices architecture, both insecure and secure.

At the conclusion of the SLR, we identified lines of research for a reliable and secure deployment of an application based on a microservices architecture. It is worth noting that in order to carry out a safe development and deployment of microservices-based systems, the processes should preferably adhere to a security model, as well as aim to provide global data protection. The lines of research identified can be briefly noted as:

- Considering the McCumber Cube [9], extending the principles of information security with the attributes of authentication, authenticity, and authorization, as a framework for building software around a microservices architecture, and thus contributing to security evaluation and auditing, as an aid to develop security policies, and determine education, training, and awareness requirements [53].

- Collecting solutions oriented to the origin of data by source, such as databases. It is critical to approach them as a security complement for the integration of a reliable deployment for the communication of systems and other elements in a microservices architecture.

Some selected studies of the SLR mention complementary solutions for the communication of microservices. These studies relate to the availability, monitoring, and optimization of resources, due to the close relationship with security for the communication of microservices. Monitoring complements some principles or information security [8], in the communication of microservices and interaction with other components and entities of the application, as monitoring aids to verify that the security attributes are fulfilling their function within the application and at the same time, it allows keeping under observation the behaviour of the parts of the application; likewise, the optimization of resources in the deployment of applications based on a microservices architecture is critical, since depending on the security implementation in the architecture, the authors mostly agree on the problem of resources, due to the security layers that developers must implement to respect the principles and characteristics of security in the communication of entities and software components. Finally, it is important to stress that availability, is crucial to deploy a complete amalgam of security in the communication of microservices.

For future work, as mentioned in section 3.3, we do not considered all the criteria from the Guidelines for performing Systematic Literature Reviews in Software Engineering proposed by [13], as we considered that the criterion for evaluating the quality of the papers to be not necessary for this study. However, a discussion including quality metrics on the studies is desirable to better assess their veracity, performance, efficiency, among other critical properties.

International Journal of Network Security & Its Applications (IJNSA) Vol.13, No.6, November 2021[35] C. Gerking, and Schubert (2019) "Component-Based Refinement and Verification of Information-Flow Security Policies for Cyber-Physical Microservice Architectures". IEEE International Conference on Software Architecture (ICSA), pp. 61-70.

[36] P. Kookarinrat, and Y. Temtanapat, (2016) "Design and implementation of a decentralized message bus for microservices". 2016 13th International Joint Conference on Computer Science and Software Engineering (JCSSE), pp. 1-6.

[37] M. Stocker, O. Zimmermann, U. Zdun, D. Lübke, and C. Pautasso, (2018) "Interface Quality Patterns: Communicating and Improving the Quality of Microservices APIs". Proceedings of the 23rd European Conference on Pattern Languages of Programs (EuroPLoP '18). Association for Computing Machinery, New York, NY, USA, Article 10, pp. 1–16.

[38] M. Pahl, and L. Donini, (2018) "Securing IoT microservices with certificates". NOMS 2018 - 2018 IEEE/IFIP Network Operations and Management Symposium, pp. 1-5.

[39] R. Tonelli, M. I. Lunesu, A. Pinna, D. Taibi, and M. Marchesi, (2019) "Implementing a Microservices System with Blockchain Smart Contracts". IEEE International Workshop on Blockchain Oriented Software Engineering (IWBOSE), pp. 22-31.

[40] C. Fetzer, (2016) "Building Critical Applications Using Microservices". IEEE Security and Privacy 14, 6, pp. 86–89.

[41] E. Aromataris, and Z. Munn, (2020) "JBI Manual for Evidence Synthesis". JBI. Available in: https://synthesismanual.jbi.global. https://doi.org/10.46658/JBIMES-20-01

[42] S. F. Johnson, and R. L. Woodgate, (2017) "Qualitative research in teen experiences living with food-induced anaphylaxis: A meta-aggregation". Journal of Advanced Nursing, Volume 73, pp. 2534-2546.

[43] K. Hannes, and A. Pearson, (2011) "Obstacles to the Implementation of Evidence-Based Practice in Belgium: a worked example of meta-aggregation". K. Hannes y C. Lockwood (Eds.), Synthesizing Qualitative Research, pp. 21-39. John Wiley & Sons.

[44] M. Howard, and S. Lipner, (2006) "The Security Development Lifecycle Process". The Security Development Lifecycle, SDL: A Process for Developing Demonstrably More Secure Software, pp. 114-116. Microsoft press.

[45] M. Cabric, (2015) "Confidentiality, Integrity, and Availability". Corporate Security Management (pp. 185-200)

[46] J. Andress, (2011) "What is Information Security?". The Basics of Information Security (Second Edition), pp. 1-16.

[47] C. Easttom, and W. Butler, (2019) "Modified McCumber Cube as a Basis for a Taxonomy of Cyber Attacks". IEEE 9th Annual Computing and Communication Workshop and Conference (CCWC), pp. 0943-0949.

[48] R. Scandariato, K. Wuyts, and W. A. Joosen, (2015) "A descriptive study of Microsoft's threat modeling technique". Requirements Engineering, 20, pp. 163–180.

[49] B. Christudas, (2019) "Microservice Performance. Practical Microservices Architectural Patterns". Apress, Berkeley, CA. pp. 279-314.

[50] S. Arnautov, B. Trach, F. Gregor, T. Knauth, A. Martin, C. Priebe, J. Lind, D. Muthukumaran, D. O'Keeffe, M. L. Stillwell, D. Goltzsche, D. Eyers, R. Kapitza, P. Pietzuch, and C. Fetzer, (2016). SCONE: secure Linux containers with Intel SGX. Proceedings of the 12th USENIX conference on Operating Systems Design and Implementation (OSDI'16). USENIX Association, USA, pp. 689–703.

[51] J. Manferdelli, T. Roeder, and F. Schneider, (2013) "The CloudProxy Tao for Trusted Computing". Technical Report UCB/EECS-2013-135. Electrical Engineering and Computer Sciences University of California at Berkeley.

[52] S. Becker, S. Dziwok, C. Gerking, C. Heinzemann, W. Schäfer, M. Meyer, and U. Pohlmann, (2014) "The MechatronicUML method: model-driven software engineering of self-adaptive mechatronic systems". Companion Proceedings of the 36th International Conference on Software Engineering (ICSE Companion 2014). Association for Computing Machinery, pp. 614–615.

[53] J. P. Myers, and S. Riela, (2008) "Taming the diversity of information assurance & security". Journal of Computing Sciences in Colleges, 23, 4, pp. 173–179.
103